\documentclass{easychair}

\usepackage{doc}
\usepackage[disable]{todonotes}

\usepackage{booktabs}
\usepackage{dirtytalk}
\usepackage{amssymb}

\title{First Experiments with Neural cvc5}

\author{
Jelle Piepenbrock\inst{1, 2}
\and
   Mikol\'a\v s Janota\inst{2}
\and
   Josef Urban\inst{2}
\and
   Jan Jakub\r uv\inst{2,3}   
}

\institute{
  Radboud University,
  Nijmegen, the Netherlands\\
  \email{jelle.piepenbrock@ru.nl}
\and
  Czech Technical University in Prague,
  Prague, Czech Republic \\
  \email{firstname.lastname@cvut.cz}\\
\and
  University of Innsbruck,
  Innsbruck, Austria
 }
\authorrunning{Piepenbrock, Janota, Urban and Jakub\r uv}

\titlerunning{First Experiments with Neural cvc5}

\begin{document}

\maketitle

\begin{abstract}
  The cvc5 solver is today one of the strongest systems for solving
  first order problems with theories but also without them. In this
  work we equip its enumeration-based instantiation with a neural network
  that guides the choice of the quantified formulas and their 
  instances. For that we develop a relatively fast graph neural
  network that repeatedly scores all available instantiation options
  with respect to the available formulas. The network runs directly on
  a CPU without the need for any special hardware. We train the neural
  guidance on a large set of proofs generated by the e-matching instantiation strategy and evaluate its performance on a
  set of previously unseen problems.
\end{abstract}

\section{Introduction}%
\label{sec:intro}
In recent years, machine learning (ML) and neural methods have been
increasingly used to guide the search procedures of automated theorem
provers (ATPs). Such methods have been so far developed for choosing
inferences in connection tableaux
systems~\cite{UrbanVS11,KaliszykU15,KaliszykUMO18,RawsonR21,ZomboriUO21}, resolution/superposition-based
systems~\cite{JakubuvU17a,DBLP:conf/cade/JakubuvCOP0U20,GoertzelCJOU21,DBLP:conf/cade/000121a}, SAT solvers~\cite{selsam2019guiding}, tactical
ITPs~\cite{GauthierKU17,DBLP:conf/icml/BansalLRSW19,DBLP:conf/lpar/BlaauwbroekUG20,GauthierKUKN21,abs-1212-3618,abs-1907-07794,abs-2401-02949} and most
recently also for the iProver~\cite{DBLP:conf/cade/Korovin08}  instantiation-based
system~\cite{ChvalovskyKPU23}. In SMT (Satisfiability Modulo Theories), ML has so far been mainly used for tasks such as portfolio and strategy optimization~\cite{10.1007/978-3-030-72013-1_16,10.1007/978-3-030-80223-3_31,BalunovicBV18}.  Previous work~\cite{janota-sat22,el2021methodes,blanchette:hal-02381430} has explored fast non-neural ML guiding methods based on decision trees and manual features. Direct neural guidance of state-of-the-art SMT
 systems such as cvc5~\cite{cvc5} and Z3~\cite{z3} however
has not been attempted yet. 

One reason for that is the large number of choices typically available
within the standard SMT procedures. Any ground term that already
exists in the current context can be used to instantiate any free
variable in the problem. While e.g.\ in the resolution/superposition
systems, only the choice of the given clause can be guided and the
rest of the work (its particular inferences with the set of processed
clauses) is computed automatically by the ATPs,\footnote{This is
  similar also in the iProver instantiation-based system with its given-clause loop.} in SMT, the
trained ML system needs to make many more and finer decisions. This is
both more fragile --- due to the many interconnected low-level decisions instead of
one high-level decision --- and also slower. The ML (and especially
neural) guidance is typically much more expensive than the standard
guiding heuristics implemented in the systems, and the more low-level and exhaustive such guidance
is, the larger the slowdown incurred by it becomes.

Despite that, there is a good motivation for experimenting with neural guidance 
for instantiation. Today's
instantiation-based systems and SMT solvers such as cvc5,
iProver and Z3 are becoming
competitive on large sets of related problems coming e.g.\ from the \emph{hammer}~\cite{hammers4qed} links between ITPs and ATPs, and even for problems that do not contain explicit theories in the SMT
sense~\cite{BlanchetteBP13,DesharnaisVBW22,GoertzelJKOPU22}. The problems that they solve are often complementary to those solved by
the state-of-the-art superposition-based systems such as E~\cite{Schulz13} and Vampire~\cite{Vampire}.
%
\vspace{2mm}
\\
\textbf{Contributions:}
In this work we develop efficient neural guidance for the enumerative
instantiation in cvc5.  We first give a brief overview of the
instantiation procedures used by cvc5 (and generally SMTs) in
Section~\ref{sec:inst}. We then design a graph neural network (GNN)
that is trained on cvc5's proofs and tightly integrated with cvc5's
proof search. This yields a neurally guided version of cvc5 that runs
reasonably fast without the need for specialized hardware, such as
GPUs. Section~\ref{sec:mlinst} explains the GNN design, its training,
collection of the training data from cvc5 and the neural instantiation
procedure. Finally in Section~\ref{sec:experiments}, we show that the
GNN-guided enumeration mode outperforms cvc5's standard enumeration by 72\%
in real (CPU) time. This is measured on previously unseen
problems extracted from the Mizar Mathematical library, after
training the GNN on many proofs obtained on a large training set.  We
also investigate the behavior of cvc5's instantiation strategies, in
terms of the number of instantiations performed in successful proof
attempts. We show that e-matching can instantiate many more times than
the enumeration strategies on our 
dataset. When we compensate for this, we arrive at an ML solver that improves on 
the enumeration mode by 109\% in real time.

\section{Proving By Instantiation}\label{sec:inst}
Quantifiers are essential in mathematics and reasoning.
Practically, all today's systems used to formalize mathematics and for
software verification are based on foundations such as first-order and
higher-order logic, set theory and type theory, which make extensive use of
quantification. Instantiation is a powerful method 
for formal reasoning with quantifiers. 
For example, the statement \say{All countries are completely in the northern hemisphere} is a quantified (false) statement, where \say{All countries} is a 
quantifier.
This statement is readily shown false by instantiating with the country Brazil.
The power of instantiation is formalized by \emph{Herbrand's theorem}~\cite{herbrand},
which states that a set $S$ of first-order clauses is unsatisfiable if and only if there is a finite set of ground instances of $S$ that is unsatisfiable.
In other words, quantifiers in false formulas can always be eliminated by a finite number of appropriate instantiations.
Herbrand's theorem further states that  it is sufficient to consider instantiations from the \emph{Herbrand universe},
which 
consists of
ground terms constructed from the symbols appearing in the
problem. 
This 
fundamental
result has been explored in automated reasoning (AR) systems since the 1950s~\cite{Davis01, DP:JACM-1960, Schulz:FLAIRS-2002,DBLP:conf/lics/GanzingerK03}.

\begin{figure}[b]
    \centering
    \includegraphics[width=0.4\textwidth]{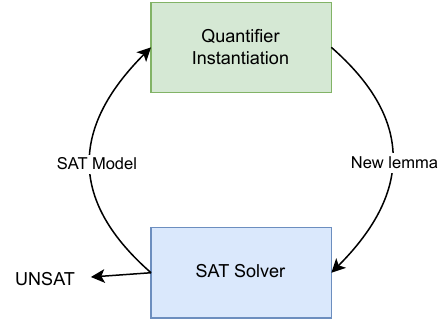}
    \caption{High-level architecture of cvc5. The techniques explained in Section~\ref{sec:inst}, as well as our neural method (Section~\ref{sec:mlinst} and on), are particular cases of the top rectangle.}%
    \label{fig:rounds}
\end{figure}

SMT solvers such as cvc5 and z3 handle quantifiers in a loop illustrated by
Figure~\ref{fig:rounds}. The loop alternates between the \emph{quantifier
module}, which provides new instantiations (called lemmas), and the
\emph{SAT solver}, which decides whether the instantiations already lead to
unsatisfiability.  In the context of this work, we refer to one iteration of
this high-level loop as a \emph{round}.

There is extensive work on techniques that calculate new instantiations. Here we
provide a concise explanation of three different quantifier instantiation
methods implemented in cvc5. While e-matching is usually seen as the standard
method (and is in fact part of the default procedure of cvc5), we start with
the enumerative instantiation method in our explanation, as it is relatively
simple and allows us to introduce concepts more gradually.
%
\vspace{2mm}
\\
\textbf{Enumerative Instantiation:}
exhaustively instantiates with  ground terms present
in the current context~\cite{reynolds2018revisiting}. In each round, it
instantiates each \emph{quantified expression}\footnote{These are
  in general formulas, however in our experiments here we restrict ourselves to
  clausal problems.} with a tuple of  ground terms---one term per variable. For a schematic overview, see Figure~\ref{fig:process_schematic}.
The default strategy of the solver is to first try the instantiations that use
terms that were created earlier in the process 
(or were already in the input problem) --- we refer to
this as the \textit{age heuristic}. For instance, for the ground part
$\{p(c)\}$ and the quantifier $\forall x.\,q(f(x))$, the solver would
instantiate by $c$ in the first round and by $f(c)$ in the second round.
There is also the option to restrict the set of terms using the
\textit{relevant domain}, but in our experiments we turn this
off. Disabling this option simplifies the integration of the machine learning component. The relevant domain option is also turned off in our machine learning experiments. When we say enumeration mode, we mean the pure enumeration
without \textit{relevant domain}. The enumeration procedure only
produces one instantiation per quantified expression per round.
%
\vspace{2mm}
\\
\textbf{E-matching:}
%
In e-matching instantiation~\cite{detlefs2005simplify,moskal2008matching}, the
solver looks for instantiations (substitutions) that yield an existing ground
term, modulo equality. Since there may be many such terms, it only considers
terms fitting a certain pattern---called \emph{trigger}. Triggers may be
provided by the user or generated by heuristics. As an example, consider the
ground facts $\{a=f(17), p(a)\}$ and the quantifier formula $\forall
x.\,\lnot p(f(x))\lor x<0$.
The trigger $p(f(x))$ would cause e-matching to instantiate $x$ with $17$
because the term $p(f(17))$ fits the trigger and it is semantically equal to
the existing term $p(a)$. This instantiation would yield the lemma
$\lnot p(f(17))\lor 17<0$, which would give a contradiction with the ground facts.
Preliminary investigations indicate that it is possible to use a similar
machine learning setup as we used here to choose the triggers, but we leave it
for future work here. Note that in contrast to the enumeration mode, e-matching may produce (potentially many) more than 
one instantiation per quantified expression per round. This difference in generative capacity becomes relevant in our experiments (Section~\ref{sec:experiments}). 
%
%
%
\vspace{2mm}
\\
\textbf{Conflict-driven Instantiation:}
In conflict-driven quantifier instantiation~\cite{reynolds-fmcad14}, the solver
looks for easily-detectable contradictions between the quantified part and the
current model of the ground part; this reasoning is done modulo equality. As an
example, consider a model $M$ that contains the facts
$\{f(a)\neq g(b),b=h(a)\}$ and there is a quantified formula $\forall
x.\,f(x)\neq g(h(x))$. Then, the solver quickly identifies that instantiating $x$
with $a$ causes a contradiction with $M$ and therefore, yielding the lemma
$f(a)\neq g(h(a))$. Adding this lemma effectively excludes $M$ from further
search. The method only looks for instantiations that guarantee a conflict with
the current model and aims to be fast and is therefore inherently incomplete. This technique is part
of the default settings of cvc5 and we can turn this off using \verb|--no-cbqi| to arrive at a solver that uses only e-matching to instantiate.
\vspace{2mm}
\\
\textbf{Term Creation \& Proliferation:}
In any of these strategies, new ground terms are created by the
instantiations that they propose. For example, when a subterm
$f(f(X))$ is instantiated with a constant $c$, the new ground term
$f(f(c))$ is created, as well as its subterm $f(c)$, if this subterm
did not already exist as a ground term in the problem. Potentially,
this can create a lot of new terms, which make the problem of choosing
terms for instantiations harder.

\section{Neural Instantiation for cvc5}
\begin{figure}[b]
    \centering
    \includegraphics[width=0.7\textwidth]{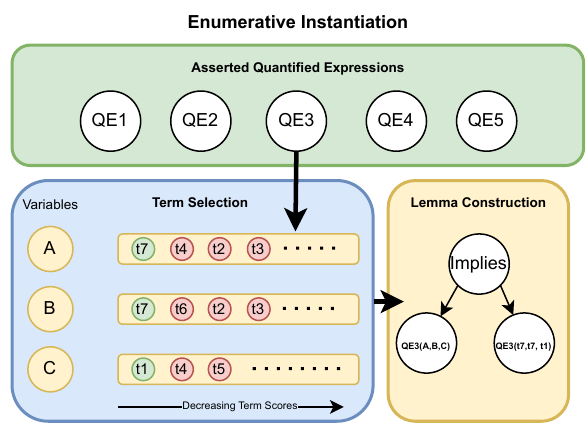}
    \caption{A schematic description of 1 instantiation within the enumerative instantiation procedure, which we heavily modify to create our neural solver. In the example a new ground lemma is created by instantiating the variables (A,B,C) in quantified expression 3 (QE3) with the ground terms (t7, t7, t1) respectively. In the default enumerative instantiation procedure, all quantified expressions are instantiated each round.}
    \label{fig:process_schematic}
\end{figure}
\label{sec:mlinst}

\textbf{Setting:}
We build our neural guidance on top of the enumerative instantiation.
This is because (i) enumeration
is the conceptually simplest of the instantiation strategies,
(ii) it
is general and complete, and (iii) 
it allows fine-grained
control over the instantiations (which can however also have the downsides mentioned in Section~\ref{sec:intro}).
That said, our current neural
guidance method is not necessarily complete --- it may be very
unfair. This is exactly the objective of training a machine learning
heuristic: we want to learn from previous proofs how we can push the
solver to be biased towards choices similar to the ones that were
previously successful. While in principle, it is possible to create
the training data for the neural network also from the enumerative instantiation
mode, we chose to use the e-matching procedure for that.
This is because e-matching is much stronger on our dataset of FOL
problems (see Section~\ref{sec:experiments}), providing us with much
more training data. This is similar to the experiments done with the E
and ENIGMA systems in~\cite{Goertzel20}, where the training data collected from the
``smart'' E strategies are used to train a guidance for a \emph{tabula
  rasa} version of E.
%
%
\todo{Figure of the neural network (can perhaps be combined with the RL setting figure.)}
%
\vspace{2mm}
\\
\textbf{GNN:} While many different neural network methods can be
used to guide automated theorem provers, a natural choice, based on
the graph representation that cvc5 uses for the proof state, is the
class of predictors called \textit{graph neural networks} (GNNs)~\cite{scarselli2008graph}. On a
high level, GNNs represent each node in a graph with a vector of
floating point numbers, and update these vectors using the vector
representations of neighbouring nodes in the graph. By using
optimization procedures, the GNN `learns' to aggregate and update the
node representations in such a way that at the end of several
iterations of this neighbourhood-based updating procedure (usually
called \textit{message passing}), the node representations contain
useful information to predict some relevant quantity. In our setting,
these relevant quantities are: (i) scores for each quantified
expression that represent whether this expression should be
instantiated and (ii) scores for each pair of variables and terms that
represent whether this particular variable should be instantiated with
a particular term. 

We implement a custom GNN using the C++ frontend of
PyTorch~\cite{paszke2019pytorch}. ML-guided ATPs often use a separate GPU
server~\cite{ChvalovskyKPU23,GoertzelJKOPU22}, to which multiple
prover processes send their requests for advice. Here, we are however interested
in a tight integration within cvc5, 
allowing the ML
component to use only one CPU thread. This also means
that no time is lost communicating and encoding/decoding between different processes and different programming languages.
\vspace{2mm}
\\
\textbf{GNN Proof State Representation:}
Each cvc5 state is represented as a graph. 
Its nodes represent cvc5 expressions. They have
a \textit{kind}, such as \textit{BOUND VARIABLE}, 
\textit{APPLY}, etc. Each of the \textit{kinds} that cvc5
recognizes internally is given a separately trainable embedding (vector in $\mathbb{R}^n$) that
serves as the initial embedding of each node before the message
passing phase (see below).
The edges between nodes are collected by modifying the DAG that cvc5 uses to represent
the state. The GNN uses a bidirectional (cyclic) version of the DAG\@. For example, if we have a term $f(c)$, we have not only an edge from $f$ to $c$, but also an edge going the other
way. We also use edge types to encode the argument ordering. For
example, in $f(c, d)$ the edge from $f$ to $c$ has a different
type than the edge from $f$ to $d$.
We recognize up to 5 different argument
positions, with the fifth used to represent all remaining positions. Each edge type has a numerical vector associated with it that is used to make it possible to distinguish argument order during the message passing procedure.
\vspace{2mm}
\\
\textbf{GNN Message Passing:}
For the message passing part of the GNN, a concatenation of the \textit{mean} and \textit{max}
aggregation of neighbourhood messages is used. We have also implemented and tested the simpler \textit{mean} 
and the more complicated \textit{attention-based} aggregation
methods. However, the \textit{mean-max}
version had the best balance of complexity and performance.\footnote{Here we consider both the
complexity of the implementation and the computational complexity of the quadratic attention mechanism.} Similarly, we tested the GNN with 2, 4,
10, 20 and 30 message passing layers, all with separate parameters. We
found that 10 layers was a reasonable trade-off between the extra
capability to fit the data and the execution speed within
cvc5. \todo{JU: point here to the evaluation of speed and accuracy and  overall performance}

The different edge types are handled by adding a trainable vector $e$ (which differs for each of the recognized edge types) to each source node in an edge, before doing the neighbourhood accumulations. This method avoids having separate weight matrices and thus message passing rounds for each edge type~\cite{schlichtkrull2018modeling}, 
which could complicate and possibly slow down the computations. In addition, these edge vectors can be seen as analogous to the \textit{positional encodings} often used in transformer architectures. The embedding update rules (for embedding size K) are as follows for a single node j with N neighbours labeled by i:
\begin{align*}
  s_t &= x_t + e_{\textrm{type}_{ij}} \\
  x_{t+1}^j &= \textrm{RELU}(W \cdot \textrm{CONCAT}(\frac{1}{N} \sum_{i=0}^N s_t^i, \max (s_t))) + x_t^j,
\end{align*}
where we compute a ``source" vector $s$ for each neighbour i depending on the type of the edge from i to j. The MAX function returns a vector that is the elementwise maximum over all the neighbourhood vectors. The matrix $s_t$ has the shape $N \times K$ and $x_{t+1}^j$ is a vector of size $K$. $W$ is a linear transformation from dimensions $2K$ to $K$. CONCAT is a concatenation function that takes two vectors of size $K$ and returns a vector of size $2K$. $\text{RELU}$ is the Rectified Linear Unit function, max(0, x), computed elementwise. The above is performed for each node and its neighbourhood, in each message passing step. Each message passing step uses its own weight matrix $W$. At the end of each message passing step, we add to each node the associated $x_t$ vector, which serves as \textit{residual connection}, a way to easily propagate information unchanged through the layers, if it is useful. In our experiments, the embedding size K was set to 64.
%
\vspace{2mm}
\\
\textbf{GNN Output:}
To decide what the solver should do, we use two different outputs of
the GNN (see also Figure \ref{fig:process_schematic}): (i) probability distributions for each of the the top-level asserted quantified
expressions, and (ii) for each variable a probability distribution
over the terms of the right type, which we interpret as a probability
that substituting this term in the variable leads to a useful
instantiation. \todo{JU: again, the following stuff should be explained in simple terms and the detailed explanation moved to an appendix - too much ML/neural details.}
For output (i), note that we output a separate prediction for each quantified expression, which we can interpret as the probability that this quantified expression will be useful for the proof.

After the message passing steps, we have embeddings
corresponding to each node in the graph. For the quantified expression
selection task, we take all the nodes corresponding to the currently
asserted top-level quantified expressions and apply a matrix
transformation (a Linear layer) of size $K \times 1$ to each one, and use a sigmoid transformation to obtain a
score between 0 and 1 for each one.  Binary
cross-entropy loss is used train the network to optimize the scores
according to the data.

For the term ranking task, we take the embedding representing
variables, and then for each variable the embeddings representing
terms of the correct type. We apply separate trainable linear transformations (of size K to K)
the term and variable embeddings and then compute dot products to
obtain a distribution of term scores for each variable. We use a
softmax transformation and cross-entropy loss to train the network to
give high scores to variable-term substitutions occurring in the data.
\vspace{2mm}
\\
\textbf{Training Data Extraction:}
\label{sec:traindat}
We have modified cvc5 to export the current proof state as a
graph. For a particular problem, we extract for each solver round
(Section~\ref{sec:inst} and Figure~\ref{fig:rounds}) the graph representation corresponding to the current proof
state. To each such graph we also assign labels that indicate which
quantified expressions and their instantiations were useful for the proof.
%
In particular, our exporter labels instantiations as the correct answer as soon as the right ground terms 
become available. We also keep track of which instantiations were already done
at a certain point in the run, so the GNN is not instructed to
repeat instantiations. When there are multiple useful instantiations
for the same quantified expression in a given proof state, we pick one at random.
This is motivated by the
\textit{enumerative instantiation} mode's default behavior, where we
only instantiate each expression once in every round. 
\vspace{2mm}
\\
\textbf{Training Details:}
For a given set of training problems, we might have many transitions for a single problem and few for another one. To balance this out, in each \textit{iteration} over the dataset, we randomly sample one of the transitions associated with the known proof for each training problem. The Adam optimizer implemented in PyTorch was used with default parameters, except for the learning rate, which was set to 0.0001.
\section{Experiments}
\label{sec:experiments}
All our experiments were run on a machine with Intel(R) Xeon(R) CPU
E5-2698 v4 @ 2.20GHz CPU, 512GB RAM and NVIDIA Tesla V100 GPUs. The GPUs 
were used only for training the GNN. The trained neural models were always run on a single CPU when used for prediction inside cvc5.
Our code and the trained GNN weights are
available from our public repository.\footnote{\url{https://github.com/JellePiepenbrock/mlcvc5-LPAR}}


\subsection{Small Dataset}

We first experimented with a small set of related problems extracted
from the Mizar library. In particular, the MPTP2078
benchmark\footnote{\url{https://github.com/JUrban/MPTP2078}}~\cite{abs-1108-3446}
has been used for several earlier AI/TP experiments~\cite{KuhlweinLTUH12-long,KaliszykU15,KaliszykUMO18}. To
make this smaller benchmark compatible with the larger benchmark we
ultimately use (see below), we update the problems there to their
version corresponding to the MML version 1147. This yields 2172
problems. These problems were split into three different sets, the
\emph{training} set (80\%), the \emph{devel}opment set (10\%) and the \emph{holdout} set
(10\%). The training set consists of problems where we are allowed to
learn from solutions that we have, the development set consists of
problems that we may tune the performance of our algorithm on and the
holdout set is a set that is not used to tune parameters.

To collect training data for our approach, we ran our modified
(Section~\ref{sec:mlinst}) version of cvc5 in \textit{only e-matching}
mode.\footnote{This means that we use the \texttt{--no-cbqi} and
  \texttt{--no-cegqi} parameters.} The states, as well as the
instantiations done were logged. These were converted into training
data using the procedure described in Section~\ref{sec:traindat}. We
end up with 814 solved problems, from which we extract 1934 training
transitions. The model was then trained for 2000 iterations
\todo{JU:  iterations are epochs? 2000 sounds excessive. JP: They're not really epochs - I take 1 transition each iteration (so not all for each proof). the 2K iterations network was better than the 1K one on the devset.)}
over the dataset.

In Table~\ref{tab:table_mptp}, we show the results of running various
versions of cvc5 for 10s on the 
development and holdout
sets.\footnote{In the evaluations, we always use 15 parallel processes, however each problem always uses a single CPU.}
The top 3 rows in the table
correspond to a binary release of cvc5. The \textit{enumeration} mode
is observed to be a lot weaker on this dataset than
\textit{e-matching}. We also show a \textit{dry run}, which is a run where we call the GNN
to compute all the scores for quantified expressions (QEs) and
term-variable combinations, 
but where we ignore those scores and simply use the
default \textit{enumeration} strategy's suggestions. This allows us to
gauge the 
slowdown caused by calling the GNN and its message passing and scoring procedures. 
While the 10s timeout can be seen as relatively short compared to the multi-minute timeouts used in competitions like SMTCOMP, it is indicative of the performance in a hammer-type setting.

We can use the scores (which are between 0 and 1 for each QE)
predicted by the GNN in different ways. Here, we experimented with two
procedures for the quantified expression scores: (i) \textit{QSampling}, where
we take the scores associated with each QE and interpret this as the
probability of using this QE in this round. A sampling procedure
decides, by drawing a random number between 0 and 1 and comparing with
the score given, whether to instantiate this QE in this round. If the
random number is higher than the score, we do not use the 
QE in this round. (ii) Threshold, where we compare the score to a
threshold number and only instantiate the QEs with scores above the
threshold. In our tuning phase (done on the development set), we found that a very low threshold\footnote{We tested the following thresholds: 
0.5, 0.4, 0.3, 0.2, 0.1, 0.01, 0.001, 0.0001 and 0.00001.} value ($0.00001$) was the best
setting for the Threshold variant. In general, false negatives (preventing a QE to be instantiated) seem to be a bigger
problem for the solver than false positives. As in premise selection, having some extra QEs instantiated does not seem to be as problematic as omitting some necessary ones.
\begin{table}[]
    \centering
\begin{tabular}{lrr}
\toprule
 & Development & Holdout \\
\midrule
Bcvc5 --- default strategies & 148 & 134 \\
Bcvc5 --- only e-matching & 129 & 115 \\
Bcvc5 --- only enumeration & 48 & 49\\
MLcvc5 --- dry run & 33 & 22 \\
MLcvc5 --- model (QSampling) & 49 & 29 \\
MLcvc5 --- model (Threshold --- 1 $\times$ 10-5) & 54 & 35 \\
\bottomrule
\end{tabular}
    \caption{Number of problems solved by 10s runs on the devel and holdout parts of the small dataset. Bcvc5 is an unchanged binary release of cvc5. MLcvc5 is our modified version. Some of the changes cause a slowdown.}
    \label{tab:table_mptp}
\end{table}

Comparing the `Bcvc5 --- only enumeration' and `MLcvc5 --- dry run' in Table~\ref{tab:table_mptp} we see
that the performance hit caused by calling the GNN is quite significant. However, we see that
both on the development and holdout sets we do improve on the performance
of the \textit{dry run}. This means the predictions of the GNN are
useful. Unfortunately, we are not able to improve on the binary
version of \textit{enumeration} on the holdout set. On the development
set, we can improve on it by a few problems. The split between training, development and holdout was done randomly to prevent a bias in the different sets. However, several of the methods seem to have worse performance on the holdout set here. A larger selection of problems could help alleviate this. While these results indicate that the GNN can be useful for guidance of cvc5 in the \textit{enumeration} mode, this training dataset might be
 too small to learn sufficiently strong GNN guidance.


\subsection{Large Dataset (MPTP1147)}
In the final evaluation we use the full MPTP 1147
benchmark, used previously in the Mizar40~\cite{KaliszykU13b} and Mizar60~\cite{JakubuvCGKOP00U23}
experiments. This dataset is more than 25x as large as the MPTP2078-induced 
subset (57917 problems in total). We use the train/devel/holdout split as defined in the
previous work~\cite{JakubuvCGKOP00U23,ChvalovskyKPU23}. Running on the training problems
with the data collection mode of cvc5 with only e-matching active
gives us 10945 solved problems, which we use to generate the training
data. The model was trained for 150 iterations on this training
dataset. After 150 epochs, the model has 82.9\% accuracy on predicting the right terms for each variable on the training problems. For the quantified expression task, which scores the QEs between 0 and 1, 70.7\% of useful quantified expressions are given a score above 0.5, while 88.3\% of the non-useful quantified expressions are given a score below 0.5. While the model did not perfectly fit the training data, it has some capacity to learn the data.  
\begin{table}[h]
    \centering
\begin{tabular}{lrr}
\toprule
 & Development & Holdout \\
\midrule
Bcvc5 --- only e-matching & 1096 & 1107 \\
Bcvc5 --- only enumeration & 183 & 195 \\
MLcvc5 --- dry run & 119 & 120 \\
MLcvc5 --- model (QSampling) & 288 & 300 \\
MLcvc5 --- model (Threshold - 1x10-5, 1 inst) & 324 & 336 \\
MLcvc5 --- model (Threshold - 1x10-5, 10 inst) & 410 & 407 \\
\bottomrule
\end{tabular}
    \caption{MML1147: Number of problems solved by 10s runs. On both the development and the holdout sets the GNN-guided enumeration mode outperforms the unguided enumeration mode. Both the development and holdout set contain 2896 problems.}
    \label{tab:mml}
\end{table}
In Table~\ref{tab:mml}, we show the development and holdout set performance of the ML-guided cvc5, along with various control experiments. We observe that the performance of the enumeration mode was improved by up to 72\% ($336/195 = 1.723$) when we use the Threshold (1 inst) variant of our network. As a sanity check, we also ran a \textit{randomized dry run} experiment on the development set, where we computed all the model scores, but used a randomly shuffled term ordering (instead of the age-based ordering for the usual dry run). This mode only solved 10 problems from the development set. These results taken together indicate that the network has learned a useful strategy from the e-matching generated training data, which it can apply in the ML enumeration mode. 
\begin{figure}[b]
    \centering
    \includegraphics[width=0.6\textwidth]{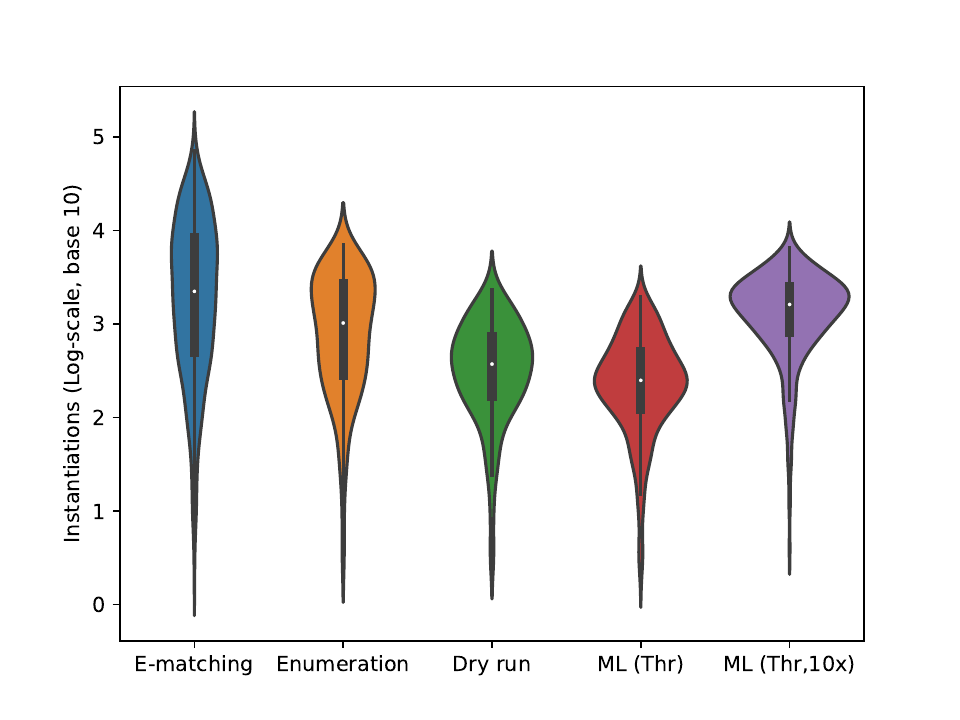}
    \caption{
    Violin plots of the number of instantiations performed in successful runs for Bcvc5 e-matching, Bcvc5 enumeration, dry run, the ML strategy with threshold 1e-5, and the ML strategy with 10x as many instantiations per quantifier per round. The white dots indicate the medians. The respective medians are 2235, 1026, 373.5, 250 and 1620.}
    \label{fig:violin}
  \end{figure}
 \vspace{2mm}
\\  
\textbf{Comparison of number of instantiations:}
While the runtime of all solvers was limited to 10s on 1 CPU, the
various versions and settings of cvc5 can vary 
in terms of
the absolute number of instantiations done within the same 
real time. The enumeration mode strives to perform at least 1
instantiation for each QE per \emph{round} (Figures~\ref{fig:rounds} and~\ref{fig:process_schematic}), and will not generally instantiate more than
once for each QE in each round. E-matching, however, is not bound by this
and will instantiate based on the number of pattern
matches (which can be high).
In Figure~\ref{fig:violin}, we show the number of instantiations done
in successful runs for five strategies. The median number
for e-matching is an order of magnitude higher
than in the ML strategy. E-matching is 
much more prolific than
enumeration, and the ML strategy is less prolific in 10s than
enumeration due to a combination of QEs that are
not used and the GNN slowdown. The number of instantiations is of course also mediated by the time spent in computing the neural network's predictions. This time varies heavily per problem and potentially per run, depending on the size of the initial problem and how much the graph grows each round (for example due to a lot of new lemmas and terms). In the successful runs on the MPTP1147 development set, there are neural network predictions that take below 40ms and ones that take more than 2400ms. 
 \vspace{2mm}
\\  
\textbf{GNN run with multiple instantiations:}
The above analysis 
indicates that some 
difference in performance is due to the
difference in the raw number of instantiations done. 
As we are
already incurring the cost of computing the GNN advice, it
might be the case that instantiation with multiple high-scoring tuples
per round, instead of only 1 per QE as the original
enumeration does, is a better use of the GNN advice. To test this, we ran
a version of the ML mode that performs up to 10 instantiations per
QE per round (see Table~\ref{tab:mml}). This led to 407 solved
holdout problems (again in 10s real time). This is a 109\% increase compared to the unmodified
enumeration mode ($407/195=2.09$).
\vspace{2mm}
\\
\textbf{Overlaps of sets:} \todo{@Josef: here. perhaps on holdout is
  most interesting} In Table~\ref{tab:setdiff}, we show the set
differences between the sets of solved problems for e-matching,
enumeration and the best-performing ML setting on the holdout set. We
observe that we can solve many problems that were not solved by the
unguided enumeration mode, but that the e-matching mode is stronger
than our method on this dataset.
\begin{table}[]
\small
    \centering
\begin{tabular}{lrrr}
\toprule
 & Bcvc5 - e-matching & Bcvc5 - enum & ML - (Thr-1x10-5, 10 inst) \\
\midrule
Bcvc5 --- e-matching & 0 & 922 & 717 \\
Bcvc5 --- enum & 10 & 0 & 34 \\
ML --- (Thr-1x10-5, 10 inst) & 17 & 246 & 0 \\
\bottomrule
\end{tabular}
    \caption{Set differences in terms of number of solved problems on holdout set, row minus column. Example: the ML solves 246 problems that the enumeration mode does not.}
    \label{tab:setdiff}
\end{table}

\section{Conclusion}
In this work, we have shown that it is possible to improve cvc5's
enumerative instantiation by 
using an 
efficient graph neural
network trained on many related problems.
Our best result is 109\% improvement in (realistically low) real time and with exactly the same hardware resources (i.e., a single CPU).
This is done here so far in a first-order clausal setting without theories, however extensions to non-clausal SMT with theories
should be mostly straightforward.



%

While e-matching largely dominates on first-order logic problems
extracted from the Mizar Mathematical Library, on problems from the
SMTLIB database, the enumeration procedure is much more
competitive~\cite{janota2021fair}, and can even outperform e-matching
on certain types of benchmarks. In principle, we can extend the
current method to SMT problems, aside from the fact that the logging
procedure that extracts training data from cvc5 runs needs to be
modified\todo{some difficulties}. This could lead to some difficulties
with tracing of instantiations, as it is not always clear how a term
came to be (as the mechanism may be `hidden' inside a theory
component).

Future work will also investigate whether a better balance between the
computation of the advice itself and the number of instantiation done
based on this advice can be found. We may be under-utilizing the
expensive advice of the GNN\@. Another direction of investigation will
be the optimization of the speed of the neural network: it may be
possible to use a much smaller neural network and still get reasonable
predictions, but much faster. We will also investigate the generalization performance of the method. For example, testing the performance on problems extracted from other systems, such as Isabelle or Coq could be insightful. In general, our work shows for the first time that efficient real-time neural guidance for SMT solvers is possible.

\section{Acknowledgements}
This work was partially supported by the European Regional Development Fund under the Czech project AI\&Reasoning no. CZ.02.1.01/0.0/0.0/15 003/0000466 (JP, JU), the European Union under the project ROBOPROX (reg.~no.~CZ.02.01.01/00/22\_008/0004590) (MJ), Amazon Research Awards (JP, JU), by the Czech MEYS under the ERC CZ project POSTMAN no.~LL1902 (JP, MJ, JU, JJ),  ERC PoC grant no.~101156734 \emph{FormalWeb3} (JJ), 
 and the EU ICT-48 2020 project TAILOR no. 952215 (JU).

\label{sect:bib}
\bibliographystyle{plain}
\bibliography{ate11}

\begin{thebibliography}{10}

\bibitem{abs-1108-3446}
Jesse Alama, Tom Heskes, Daniel K\"{u}hlwein, Evgeni Tsivtsivadze, and Josef
  Urban.
\newblock Premise selection for mathematics by corpus analysis and kernel
  methods.
\newblock {\em J. Autom. Reasoning}, 52(2):191--213, 2014.

\bibitem{BalunovicBV18}
Mislav Balunovic, Pavol Bielik, and Martin~T. Vechev.
\newblock Learning to solve {SMT} formulas.
\newblock In {\em NeurIPS}, pages 10338--10349, 2018.

\bibitem{DBLP:conf/icml/BansalLRSW19}
Kshitij Bansal, Sarah~M. Loos, Markus~N. Rabe, Christian Szegedy, and Stewart
  Wilcox.
\newblock {HOList}: An environment for machine learning of higher order logic
  theorem proving.
\newblock In {\em {ICML}}, volume~97 of {\em Proceedings of Machine Learning
  Research}, pages 454--463. {PMLR}, 2019.

\bibitem{cvc5}
Haniel Barbosa, Clark~W. Barrett, Martin Brain, Gereon Kremer, Hanna Lachnitt,
  Makai Mann, Abdalrhman Mohamed, Mudathir Mohamed, Aina Niemetz, Andres
  N{\"{o}}tzli, Alex Ozdemir, Mathias Preiner, Andrew Reynolds, Ying Sheng,
  Cesare Tinelli, and Yoni Zohar.
\newblock cvc5: {A} versatile and industrial-strength {SMT} solver.
\newblock In {\em {TACAS} {(1)}}, volume 13243. Springer, 2022.

\bibitem{DBLP:conf/lpar/BlaauwbroekUG20}
Lasse Blaauwbroek, Josef Urban, and Herman Geuvers.
\newblock Tactic learning and proving for the {Coq} proof assistant.
\newblock In {\em {LPAR}}, volume~73 of {\em EPiC Series in Computing}, pages
  138--150. EasyChair, 2020.

\bibitem{BlanchetteBP13}
Jasmin~Christian Blanchette, Sascha B{\"{o}}hme, and Lawrence~C. Paulson.
\newblock Extending sledgehammer with {SMT} solvers.
\newblock {\em J. Autom. Reason.}, 51(1):109--128, 2013.

\bibitem{hammers4qed}
Jasmin~Christian Blanchette, Cezary Kaliszyk, Lawrence~C. Paulson, and Josef
  Urban.
\newblock Hammering towards {QED}.
\newblock {\em J. Formalized Reasoning}, 9(1):101--148, 2016.

\bibitem{blanchette:hal-02381430}
Jasmin~Christian Blanchette, Daniel~El Ouraoui, Pascal Fontaine, and Cezary
  Kaliszyk.
\newblock {Machine Learning for Instance Selection in SMT Solving}.
\newblock In {\em {AITP 2019 - 4th Conference on Artificial Intelligence and
  Theorem Proving}}, Obergurgl, Austria, April 2019.

\bibitem{ChvalovskyKPU23}
Karel Chvalovsk{\'{y}}, Konstantin Korovin, Jelle Piepenbrock, and Josef Urban.
\newblock Guiding an instantiation prover with graph neural networks.
\newblock In {\em {LPAR}}, volume~94 of {\em EPiC Series in Computing}, pages
  112--123. EasyChair, 2023.

\bibitem{DP:JACM-1960}
M.~Davis and H.~Putnam.
\newblock {A Computing Procedure for Quantification Theory}.
\newblock {\em Journal of the ACM}, 7(1):215--215, 1960.

\bibitem{Davis01}
Martin Davis.
\newblock The early history of automated deduction.
\newblock In {\em Handbook of Automated Reasoning}, pages 3--15. Elsevier and
  {MIT} Press, 2001.

\bibitem{z3}
Leonardo~Mendon\c{c}a de~Moura and Nikolaj Bj{\o}rner.
\newblock {Z3: An Efficient SMT Solver}.
\newblock In C.~R. Ramakrishnan and Jakob Rehof, editors, {\em TACAS}, volume
  4963 of {\em LNCS}, pages 337--340. Springer, 2008.

\bibitem{DesharnaisVBW22}
Martin Desharnais, Petar Vukmirovic, Jasmin Blanchette, and Makarius Wenzel.
\newblock Seventeen provers under the hammer.
\newblock In {\em {ITP}}, volume 237 of {\em LIPIcs}, pages 8:1--8:18. Schloss
  Dagstuhl - Leibniz-Zentrum f{\"{u}}r Informatik, 2022.

\bibitem{detlefs2005simplify}
David Detlefs, Greg Nelson, and James~B Saxe.
\newblock Simplify: a theorem prover for program checking.
\newblock {\em Journal of the ACM (JACM)}, 52(3):365--473, 2005.

\bibitem{el2021methodes}
Daniel El~Ouraoui.
\newblock {\em M{\'e}thodes pour le raisonnement d'ordre sup{\'e}rieur dans
  SMT, Chapter 5.}
\newblock PhD thesis, Universit{\'e} de Lorraine, 2021.

\bibitem{DBLP:conf/lics/GanzingerK03}
Harald Ganzinger and Konstantin Korovin.
\newblock New directions in instantiation-based theorem proving.
\newblock In {\em 18th {IEEE} Symposium on Logic in Computer Science {(LICS}
  2003), 22-25 June 2003, Ottawa, Canada, Proceedings}, pages 55--64. {IEEE}
  Computer Society, 2003.

\bibitem{GauthierKU17}
Thibault Gauthier, Cezary Kaliszyk, and Josef Urban.
\newblock {TacticToe}: Learning to reason with {HOL4} tactics.
\newblock In {\em LPAR-21, 21st International Conference on Logic for
  Programming, Artificial Intelligence and Reasoning, Maun, Botswana, May 7-12,
  2017}, pages 125--143, 2017.

\bibitem{GauthierKUKN21}
Thibault Gauthier, Cezary Kaliszyk, Josef Urban, Ramana Kumar, and Michael
  Norrish.
\newblock Tactictoe: Learning to prove with tactics.
\newblock {\em J. Autom. Reason.}, 65(2):257--286, 2021.

\bibitem{Goertzel20}
Zarathustra~Amadeus Goertzel.
\newblock Make {E} smart again (short paper).
\newblock In {\em {IJCAR} {(2)}}, volume 12167 of {\em Lecture Notes in
  Computer Science}, pages 408--415. Springer, 2020.

\bibitem{GoertzelCJOU21}
Zarathustra~Amadeus Goertzel, Karel Chvalovsk{\'{y}}, Jan Jakubuv, Miroslav
  Ols{\'{a}}k, and Josef Urban.
\newblock Fast and slow enigmas and parental guidance.
\newblock In Boris Konev and Giles Reger, editors, {\em Frontiers of Combining
  Systems - 13th International Symposium, FroCoS 2021, Birmingham, UK,
  September 8-10, 2021, Proceedings}, volume 12941 of {\em Lecture Notes in
  Computer Science}, pages 173--191. Springer, 2021.

\bibitem{GoertzelJKOPU22}
Zarathustra~Amadeus Goertzel, Jan Jakub\r{u}v, Cezary Kaliszyk, Miroslav
  Ols{\'{a}}k, Jelle Piepenbrock, and Josef Urban.
\newblock The {I}sabelle {ENIGMA}.
\newblock In {\em {ITP}}, volume 237 of {\em LIPIcs}, pages 16:1--16:21.
  Schloss Dagstuhl - Leibniz-Zentrum f{\"{u}}r Informatik, 2022.

\bibitem{herbrand}
Jacques Herbrand.
\newblock {\em Recherches sur la th\'eorie de la d\'emonstration}.
\newblock Doctorat d'\'etat, La Facult\'e des Sciences de Paris, 1930.

\bibitem{DBLP:conf/cade/JakubuvCOP0U20}
Jan Jakub\r{u}v, Karel Chvalovsk{\'{y}}, Miroslav Ols{\'{a}}k, Bartosz
  Piotrowski, Martin Suda, and Josef Urban.
\newblock {ENIGMA} anonymous: Symbol-independent inference guiding machine
  (system description).
\newblock In Nicolas Peltier and Viorica Sofronie{-}Stokkermans, editors, {\em
  Automated Reasoning - 10th International Joint Conference, {IJCAR} 2020,
  Paris, France, July 1-4, 2020, Proceedings, Part {II}}, volume 12167 of {\em
  Lecture Notes in Computer Science}, pages 448--463. Springer, 2020.

\bibitem{JakubuvU17a}
Jan Jakub\r{u}v and Josef Urban.
\newblock {ENIGMA:} efficient learning-based inference guiding machine.
\newblock In Herman Geuvers, Matthew England, Osman Hasan, Florian Rabe, and
  Olaf Teschke, editors, {\em Intelligent Computer Mathematics - 10th
  International Conference, {CICM} 2017, Edinburgh, UK, July 17-21, 2017,
  Proceedings}, volume 10383 of {\em Lecture Notes in Computer Science}, pages
  292--302. Springer, 2017.

\bibitem{JakubuvCGKOP00U23}
Jan Jakubuv, Karel Chvalovsk{\'{y}}, Zarathustra~Amadeus Goertzel, Cezary
  Kaliszyk, Mirek Ols{\'{a}}k, Bartosz Piotrowski, Stephan Schulz, Martin Suda,
  and Josef Urban.
\newblock {MizAR} 60 for {Mizar} 50.
\newblock In {\em {ITP}}, volume 268 of {\em LIPIcs}, pages 19:1--19:22.
  Schloss Dagstuhl - Leibniz-Zentrum f{\"{u}}r Informatik, 2023.

\bibitem{janota2021fair}
Mikol{\'a}{\v{s}} Janota, Haniel Barbosa, Pascal Fontaine, and Andrew Reynolds.
\newblock Fair and adventurous enumeration of quantifier instantiations.
\newblock In {\em 2021 Formal Methods in Computer Aided Design (FMCAD)}, pages
  256--260. IEEE, 2021.

\bibitem{KaliszykU15}
Cezary Kaliszyk and Josef Urban.
\newblock {FEMaLeCoP}: Fairly efficient machine learning connection prover.
\newblock In Martin Davis, Ansgar Fehnker, Annabelle McIver, and Andrei
  Voronkov, editors, {\em Logic for Programming, Artificial Intelligence, and
  Reasoning - 20th International Conference, {LPAR-20} 2015, Suva, Fiji,
  November 24-28, 2015, Proceedings}, volume 9450 of {\em Lecture Notes in
  Computer Science}, pages 88--96. Springer, 2015.

\bibitem{KaliszykU13b}
Cezary Kaliszyk and Josef Urban.
\newblock {MizAR 40 for Mizar 40}.
\newblock {\em J. Autom. Reasoning}, 55(3):245--256, 2015.

\bibitem{KaliszykUMO18}
Cezary Kaliszyk, Josef Urban, Henryk Michalewski, and Miroslav Ols{\'{a}}k.
\newblock Reinforcement learning of theorem proving.
\newblock In {\em Advances in Neural Information Processing Systems 31: Annual
  Conference on Neural Information Processing Systems 2018, NeurIPS 2018, 3-8
  December 2018, Montr{\'{e}}al, Canada.}, pages 8836--8847, 2018.

\bibitem{abs-1212-3618}
Ekaterina Komendantskaya, J{\'{o}}nathan Heras, and Gudmund Grov.
\newblock Machine learning in {Proof General}: Interfacing interfaces.
\newblock In {\em {UITP}}, volume 118 of {\em {EPTCS}}, pages 15--41, 2012.

\bibitem{DBLP:conf/cade/Korovin08}
Konstantin Korovin.
\newblock {iProver} - an instantiation-based theorem prover for first-order
  logic (system description).
\newblock In Alessandro Armando, Peter Baumgartner, and Gilles Dowek, editors,
  {\em Automated Reasoning, 4th International Joint Conference, {IJCAR} 2008,
  Sydney, Australia, August 12-15, 2008, Proceedings}, volume 5195 of {\em
  Lecture Notes in Computer Science}, pages 292--298. Springer, 2008.

\bibitem{Vampire}
Laura Kov{\'a}cs and Andrei Voronkov.
\newblock First-order theorem proving and {V}ampire.
\newblock In Natasha Sharygina and Helmut Veith, editors, {\em CAV}, volume
  8044 of {\em LNCS}, pages 1--35. Springer, 2013.

\bibitem{KuhlweinLTUH12-long}
Daniel K{\"u}hlwein, Twan van Laarhoven, Evgeni Tsivtsivadze, Josef Urban, and
  Tom Heskes.
\newblock Overview and evaluation of premise selection techniques for large
  theory mathematics.
\newblock In Bernhard Gramlich, Dale Miller, and Uli Sattler, editors, {\em
  IJCAR}, volume 7364 of {\em LNCS}, pages 378--392. Springer, 2012.

\bibitem{moskal2008matching}
Micha{\l} Moskal, Jakub {\L}opusza{\'n}ski, and Joseph~R Kiniry.
\newblock E-matching for fun and profit.
\newblock {\em Electronic Notes in Theoretical Computer Science},
  198(2):19--35, 2008.

\bibitem{paszke2019pytorch}
Adam Paszke, Sam Gross, Francisco Massa, Adam Lerer, James Bradbury, Gregory
  Chanan, Trevor Killeen, Zeming Lin, Natalia Gimelshein, Luca Antiga, et~al.
\newblock Pytorch: An imperative style, high-performance deep learning library.
\newblock {\em Advances in neural information processing systems}, 32, 2019.

\bibitem{10.1007/978-3-030-80223-3_31}
Nikhil Pimpalkhare, Federico Mora, Elizabeth Polgreen, and Sanjit~A. Seshia.
\newblock {MedleySolver}: Online {SMT} algorithm selection.
\newblock In Chu-Min Li and Felip Many{\`a}, editors, {\em Theory and
  Applications of Satisfiability Testing -- SAT 2021}, pages 453--470, Cham,
  2021. Springer International Publishing.

\bibitem{RawsonR21}
Michael Rawson and Giles Reger.
\newblock {lazyCoP}: Lazy paramodulation meets neurally guided search.
\newblock In {\em {TABLEAUX}}, volume 12842 of {\em Lecture Notes in Computer
  Science}, pages 187--199. Springer, 2021.

\bibitem{reynolds2018revisiting}
Andrew Reynolds, Haniel Barbosa, and Pascal Fontaine.
\newblock Revisiting enumerative instantiation.
\newblock In {\em Tools and Algorithms for the Construction and Analysis of
  Systems: 24th International Conference, TACAS 2018, Held as Part of the
  European Joint Conferences on Theory and Practice of Software, ETAPS 2018,
  Thessaloniki, Greece, April 14-20, 2018, Proceedings, Part II 24}, pages
  112--131. Springer, 2018.

\bibitem{reynolds-fmcad14}
Andrew Reynolds, Cesare Tinelli, and Leonardo~Mendon{\c{c}}a de~Moura.
\newblock Finding conflicting instances of quantified formulas in {SMT}.
\newblock In {\em Formal Methods in Computer-Aided Design, {FMCAD} 2014,
  Lausanne, Switzerland, October 21-24, 2014}, pages 195--202. {IEEE}, 2014.

\bibitem{abs-2401-02949}
Jason Rute, Miroslav Ols{\'{a}}k, Lasse Blaauwbroek, Fidel Ivan~Schaposnik
  Massolo, Jelle Piepenbrock, and Vasily Pestun.
\newblock Graph2tac: Learning hierarchical representations of math concepts in
  theorem proving.
\newblock {\em CoRR}, abs/2401.02949, 2024.

\bibitem{janota-sat22}
Mikol{\'{a}}\v s~Janota, Jelle Piepenbrock, and Bartosz Piotrowski.
\newblock Towards learning quantifier instantiation in {SMT}.
\newblock In Kuldeep~S. Meel and Ofer Strichman, editors, {\em 25th
  International Conference on Theory and Applications of Satisfiability
  Testing, {SAT} 2022, August 2-5, 2022, Haifa, Israel}, volume 236 of {\em
  LIPIcs}, pages 7:1--7:18. Schloss Dagstuhl - Leibniz-Zentrum f{\"{u}}r
  Informatik, 2022.

\bibitem{abs-1907-07794}
Alex Sanchez{-}Stern, Yousef Alhessi, Lawrence~K. Saul, and Sorin Lerner.
\newblock Generating correctness proofs with neural networks.
\newblock {\em CoRR}, abs/1907.07794, 2019.

\bibitem{scarselli2008graph}
Franco Scarselli, Marco Gori, Ah~Chung Tsoi, Markus Hagenbuchner, and Gabriele
  Monfardini.
\newblock The graph neural network model.
\newblock {\em IEEE transactions on neural networks}, 20(1):61--80, 2008.

\bibitem{schlichtkrull2018modeling}
Michael Schlichtkrull, Thomas~N Kipf, Peter Bloem, Rianne Van Den~Berg, Ivan
  Titov, and Max Welling.
\newblock Modeling relational data with graph convolutional networks.
\newblock In {\em The Semantic Web: 15th International Conference, ESWC 2018,
  Heraklion, Crete, Greece, June 3--7, 2018, Proceedings 15}, pages 593--607.
  Springer, 2018.

\bibitem{Schulz:FLAIRS-2002}
S.~Schulz.
\newblock {A Comparison of Different Techniques for Grounding
  Near-Propositional CNF Formulae}.
\newblock In S.~Haller and G.~Simmons, editors, {\em Proc.\ of the 15th FLAIRS,
  Pensacola}, pages 72--76. AAAI Press, 2002.

\bibitem{Schulz13}
Stephan Schulz.
\newblock System description: {E} 1.8.
\newblock In Kenneth~L. McMillan, Aart Middeldorp, and Andrei Voronkov,
  editors, {\em LPAR}, volume 8312 of {\em LNCS}, pages 735--743. Springer,
  2013.

\bibitem{10.1007/978-3-030-72013-1_16}
Joseph Scott, Aina Niemetz, Mathias Preiner, Saeed Nejati, and Vijay Ganesh.
\newblock {MachSMT}: A machine learning-based algorithm selector for {SMT}
  solvers.
\newblock In Jan~Friso Groote and Kim~Guldstrand Larsen, editors, {\em Tools
  and Algorithms for the Construction and Analysis of Systems}, pages 303--325,
  Cham, 2021. Springer International Publishing.

\bibitem{selsam2019guiding}
Daniel Selsam and Nikolaj Bj{\o}rner.
\newblock Guiding high-performance sat solvers with unsat-core predictions.
\newblock In {\em Theory and Applications of Satisfiability Testing--SAT 2019:
  22nd International Conference, SAT 2019, Lisbon, Portugal, July 9--12, 2019,
  Proceedings 22}, pages 336--353. Springer, 2019.

\bibitem{DBLP:conf/cade/000121a}
Martin Suda.
\newblock Improving enigma-style clause selection while learning from history.
\newblock In {\em {CADE}}, volume 12699 of {\em Lecture Notes in Computer
  Science}, pages 543--561. Springer, 2021.

\bibitem{UrbanVS11}
Josef Urban, Ji\v{r}\'{\i} Vysko\v{c}il, and Petr \v{S}t\v{e}p{\'a}nek.
\newblock {MaLeCoP}: Machine learning connection prover.
\newblock In Kai Br{\"u}nnler and George Metcalfe, editors, {\em TABLEAUX},
  volume 6793 of {\em LNCS}, pages 263--277. Springer, 2011.

\bibitem{ZomboriUO21}
Zsolt Zombori, Josef Urban, and Miroslav Ols{\'{a}}k.
\newblock The role of entropy in guiding a connection prover.
\newblock In {\em {TABLEAUX}}, volume 12842 of {\em Lecture Notes in Computer
  Science}, pages 218--235. Springer, 2021.

\end{thebibliography}

\end{document}